\documentclass[prd,nofootinbib,nobibnotes,showpacs,superscriptaddress]
              {revtex4}
\usepackage{graphicx}
\usepackage{amssymb}

\newcommand{\ga}{\gamma}
\newcommand{\Ga}{\Gamma}

\newcommand{\be}{\begin{equation}}
\newcommand{\ee}{\end{equation}}

\newcommand{\bea}{\begin{eqnarray}}
\newcommand{\eea}{\end{eqnarray}}
\newcommand{\bean}{\begin{eqnarray*}}
\newcommand{\eean}{\end{eqnarray*}}

\begin{document}
\draft
\title{Effects of Cosmological
Magnetic Helicity on the Cosmic Microwave Background}

\author{Tina Kahniashvili}
\email{tinatin@phys.ksu.edu}
\affiliation{Department of Physics, Kansas State University,
116 Cardwell Hall, Manhattan, KS 66506, USA}
\affiliation{
Center for Plasma Astrophysics, Abastumani Astrophysical Observatory,
2A Kazbegi Ave, GE-0160 Tbilisi, Georgia}
\author{Bharat Ratra}
\email{ratra@phys.ksu.edu}
\affiliation{Department of Physics, Kansas State University,
116 Cardwell Hall, Manhattan, KS 66506, USA}

\date{March 2005 \hspace{0.3truecm} KSUPT-05/2}
\begin{abstract}
Cosmological  magnetic fields induce temperature and polarization
fluctuations in the cosmic
microwave background (CMB) radiation. A cosmological
 magnetic field with current amplitude of order $10^{-9}$ G
is detectable via observations of CMB anisotropies. This magnetic
field (with or without helicity) generates vector perturbations
through vortical motions of the primordial plasma. This paper
shows that magnetic field helicity induces parity-odd cross
correlations between CMB temperature and $B$-polarization
fluctuations and between $E$- and $B$-polarization fluctuations, correlations
which are zero for fields with no helicity (or for any
parity-invariant source). Helical fields also contribute to
parity-even temperature and polarization anisotropies, cancelling part of  
 the contribution from the symmetric
component of the magnetic field. We give analytic approximations
for all CMB temperature and polarization anisotropy vector power spectra due
to helical magnetic fields. These power spectra offer a method for
detecting cosmological helical magnetic fields, particularly when combined
with Faraday rotation measurements which are insensitive to
helicity.

\end{abstract}
\maketitle

\section{Introduction}
\label{sec:intro}

The linear theory for the evolution of small inhomogeneities  in
the standard  spatially homogeneous and isotropic cosmological model,
developed by Lifshitz and others \cite{lifshitz}, shows that 
vector perturbations\footnote{Vector perturbations 
 are also known as
transverse peculiar velocity or vorticity perturbations.} 
 decay as the Universe
expands.\footnote{This is for adiabatic initial conditions, which are 
consistent with observations. However, a non-decaying vector perturbation 
exists if there is an initial 
 large-scale photon-baryon fluid vorticity compensated by a 
neutrino vorticity such that the total large-scale vorticity in 
 relativistic components vanishes \cite{rehnan}. 
Such a mode can be excited only after neutrino 
decoupling \cite{rehnan}.} This is  why vector perturbations  
are usually   neglected when computing CMB
fluctuations. On the other hand,
the presence a cosmological magnetic field
\cite{adams,durrer98,subramanian98b,mack02,lewis04b,review2} 
  alters this situation. In this paper we
focus on vector perturbations induced by a cosmological magnetic
field, existing since 
 radiation-matter  equality
or earlier
(for a specific inflation model see
\cite{review3}). Both an  homogeneous magnetic field,
as well as the more realistic stochastic one, induce
transverse MHD (Alfv\'en) waves \cite {adams}.
Non-decaying cosmological Alfv\'en waves
result in CMB temperature
and polarization anisotropies
\cite{durrer98,subramanian98b,mack02,lewis04b,pogosian02,gang04}.\footnote{In addition to the vector mode, a
cosmological magnetic field
generates CMB fluctuations through the scalar and tensor modes
as  described in
Refs.~\cite{adams,adams96,jedamzik98,mack02,caprini03}.}

A cosmological  magnetic field could
have helicity
\cite{cornwall97}. Magnetic field
helicity plays an important role in the MHD dynamo,
used in some models of galactic magnetic field amplification
\cite{jedamzik04}. Unlike
gravitational waves  which are
damped on scales below the Hubble radius at decoupling
 (the
corresponding damping multipole number $l\sim 100$),
Alfv\'en waves survive down to smaller damping scales ($l \sim 2000$)
\cite{lewis04b}, thus a cosmological magnetic field can affect 
 small-scale CMB fluctuations. From this point of view, the vector
mode is more relevant for constraining a cosmological magnetic field
 from CMB fluctuation measurements.

In this paper we present analytic expressions for all CMB
fluctuation vector power spectra that arise from a helical
cosmological magnetic field. In particular, we also
compute how magnetic helicity
affects  parity-even CMB fluctuation power spectra. We propose
a scheme to constrain cosmological magnetic helicity from CMB
temperature and polarization anisotropy observations. The
symmetric part of the magnetic field spectrum can be
reconstructed from  measurements of the rotation of the CMB polarization
 plane as a consequence
of the  Faraday effect
\cite{kosowsky96}; this is because
magnetic helicity does not contribute to the Faraday rotation
effect
 \cite{ensslin03,campanelli04,far}.
On the other hand, the helical part of the magnetic field
spectrum induces parity-odd cross correlations between
temperature and $B$-polarization anisotropies,
 and between $E$- and $B$-polarization anisotropies
\cite{pogosian02,caprini03}; such cross correlations
are not induced by the Faraday effect \cite{far}.\footnote{This is not
  true for a homogeneous
magnetic field \cite{scoccola04}.}

For our computations we use the formalism  of Ref.~\cite{mack02},
extending it to account for magnetic field helicity. To compute
CMB temperature and polarization anisotropy power spectra we use
the total angular momentum method of Ref.~ \cite{hu97}. Our
results are obtained using analytic approximations and are valid
for $l<500$. We present results in terms of a ratio between CMB
fluctuation contributions from  the symmetric and   helical parts
of the magnetic field power spectrum.

The rest of the paper is organized as follows. In the next
section we model the helical magnetic field source term for
vorticity (Alfv\'en) waves. In Sections III and IV we present
analytic approximations for the  parity-even and parity-odd CMB
fluctuation power spectra contributions induced by magnetic
helicity. In Section V we discuss our results and conclude. In
Appendixes A and B we present some details of the computation
including analytic approximations used for some of the
integrations.

\section{Magnetic field induced vector perturbations}
\subsection{Magnetic field source term}
\label{sec:magnetic field}
We assume the existence of a cosmological magnetic field generated during or
prior to the radiation-dominated epoch, with the energy density of the
 field a first-order perturbation to the standard
Friedmann-Lema\^\i tre-Robertson-Walker
homogeneous cosmological spacetime model.
Neglecting fluid back-reaction onto the magnetic field,
the spatial and temporal dependence of the
field separates,
${\mathbf B}(t,{\mathbf x})={\mathbf B}({\mathbf x})/a^2$; here $a$ is the
 cosmological scale factor.
As a phenomenological normalization of the magnetic field,
we smooth the field on a comoving length $\lambda$ with a
Gaussian smoothing kernel $\propto \mbox{exp}[-x^2/\lambda^2]$
 to obtain the smoothed
magnetic field 
with average value of squared magnetic field
${B_\lambda}^2 \equiv \langle {\mathbf B}({\mathbf x})
\cdot {\mathbf B}({\mathbf x})\rangle |_\lambda$ and
magnetic helicity
${H_\lambda}^2 \equiv \lambda
| \langle{\mathbf B}({\mathbf x}) \cdot
[{\mathbf \nabla} \times {\mathbf B} ({\mathbf x})] \rangle|_\lambda$.
See Ref.~\cite{far} for a more detailed discussion.

We also assume that the primordial plasma is a perfect conductor on
all
scales larger than the Silk damping wavelength
$\lambda_S$ (the thickness of the last scattering surface)
set by photon and neutrino diffusion.
We model magnetic field  damping by an ultraviolet cut-off wavenumber
$k_D=2\pi/\lambda_D$ \cite{jedamzik98,mack02},
\be
\left({k_D \over {\rm Mpc}^{-1}}\right)^{n_B + 5} \approx 2.9\times 10^4
  \left({B_\lambda\over 10^{-9}\,{\rm G}}\right)^{-2}
  \left({k_\lambda\over {\rm Mpc}^{-1}}\right)^{n_B + 3} h.
\label{kd}
\ee
Here $n_B$ is the spectral index of the symmetric part of the
 magnetic field power spectrum (see Eq.~(\ref{energy-spectrum-H}) below),
 $h$ is the Hubble constant in units of
$100$~km sec${}^{-1}$ Mpc${}^{-1}$,
$k_\lambda = 2\pi/\lambda$ is the smoothing wavenumber,
 and $\lambda_D \ll \lambda_S$.
This assumes that magnetic field damping is due to the damping of
Alfv\'en waves
from photon viscosity.

Assuming that the stochastic magnetic field is Gaussianly distributed, and
accounting for the possible helicity of the field,
the magnetic field spectrum in wavenumber space is
\cite{pogosian02}, 
\begin{equation}
\langle B^\star_i({\mathbf k})B_j({\mathbf k'})\rangle
=(2\pi)^3 \delta^{(3)}
({\mathbf k}-{\mathbf k'}) [P_{ij}({\mathbf{\hat k}}) P_B(k)  +
i \epsilon_{ijl} \hat{k}_l P_H(k)].
\label{spectrum}
\end{equation}
Here $P_{ij}({\mathbf{\hat k}})\equiv\delta_{ij}-\hat{k}_i\hat{k}_j$
is the transverse plane projector with unit wavenumber components
$\hat{k}_i=k_i/k$, $\epsilon_{ijl}$ is the antisymmetric tensor, and 
 $\delta^{(3)}({\mathbf k}-{\mathbf k'})$ is the Dirac delta function.
We use
\be
   B_j({\mathbf k}) = \int d^3\!x \,
   e^{i{\mathbf k}\cdot {\mathbf x}} B_j({\mathbf x}),~~~~~~~~~~~
   B_j({\mathbf x}) = \int {d^3\!k \over (2\pi)^3}
   e^{-i{\mathbf k}\cdot {\mathbf x}} B_j({\mathbf k}),\ee
when Fourier transforming between real and wavenumber spaces; we
assume flat spatial hypersurfaces.
$P_B(k)$ and $P_H(k)$ are the symmetric and
helical parts of the magnetic field power
spectrum, assumed to be simple power laws on large scales,
\begin{equation}
P_B(k) \equiv P_{B0}k^{n_B}=
\frac{2\pi^2 \lambda^3 B^2_\lambda}{\Gamma(n_B/2+3/2)}
(\lambda k)^{n_B},~~~~~~~~~~
P_H(k) \equiv P_{H0}k^{n_H}=
\frac{ 2\pi^2 \lambda^3 H^2_\lambda}{\Gamma(n_H/2+2)}
(\lambda k)^{n_H},\qquad k<k_D,
\label{energy-spectrum-H}
\end{equation}
and vanishing on small scales when $k>k_D$. Here $\Gamma$ is the
Euler Gamma function. These power spectra are generically
constrained by $P_B(k)\geq |P_H(k)|$ \cite{durrer03,jedamzik04},
which implies $n_H> n_B$ \cite{durrer03,caprini03}.
 In addition, finiteness of the magnetic
field energy density requires $n_B > -3$ (to prevent an infrared
divergence of magnetic field energy density).  Finiteness of the magnetic
field average helicity requires $n_H > -4$; this is  automatically
satisfied as a consequence of $n_H> n_B >-3$.

To obtain the magnetic field source  term in the transverse
peculiar velocity perturbation equation of motion we need to
extract the transverse vector part of the magnetic field
stress-energy tensor $\tau_{ij}({\mathbf{k}})$. This is done through $
\Pi_{ij}({\mathbf{k}})=(P_{ib}({\mathbf{\hat k}})
\hat{k}_j+P_{jb}({\mathbf{\hat k}})\hat{k}_i)\hat{k}_a
\tau_{ab}({\mathbf{k}}) $, and the $\Pi_{ij}$ tensor is related to the
vector (divergenceless and transverse) part of the Lorentz force
$L_i^{(V)}({\mathbf{k}}) = k_j \Pi_{ij}({\mathbf{k}})
 =P_{ib}({\mathbf{\hat k}}) {k}_a
\tau_{ab}({\mathbf{k}}) $ (Eq.~(2.16) of Ref.~\cite{mack02}). For
the normalized Lorentz force vector $\Pi_i \equiv L_i^{(V)}/k$,
the general spectrum in wavenumber space is similar to
Eq.~(\ref{spectrum});  that is,\footnote{ We thank A.~Lewis (private
communication 2004 and Ref.~\cite{lewis04b}) for pointing out a
missing $(2\pi)^3$ factor in the expression for
$\langle\Pi_i^*({\mathbf k})\Pi_j({\mathbf k^\prime})\rangle$
given in Ref.~\cite{mack02}. In what follows we  use other
results from Ref.~\cite{mack02} corrected for a similar
missing factor.}
\begin{equation}
\langle\Pi^\star_i({\mathbf k})\Pi_j({\mathbf k'})\rangle
= (2\pi)^3 \delta^{(3)}(\mathbf {k-k'})
\left[ P_{ij} f(k) + i
\epsilon_{ijq}  {\hat k}_q
g(k)\right],
\label{source-vector1}
\end{equation}
where $f(k)$ and $g(k)$ represent the symmetric and helical
parts of the vector Lorentz force power spectrum.
These functions are related to the magnetic
field stress-energy tensor spectrum through,
\begin{eqnarray}
(2\pi)^3 \delta^{(3)} ({\bf k}- {\bf k'}) f(k) &=&
\frac{1}{4}[
P_{aj} ({\mathbf {\hat k}}) {\hat k}_i 
P_{am} ({\mathbf {\hat k}^\prime}) {\hat k}^\prime_l +
P_{ai} ({\mathbf {\hat k}}) {\hat k}_j 
P_{al} ({\mathbf {\hat k}^\prime}){\hat k}^\prime_m
]
\langle \tau^\star_{ij}({\mathbf k}) \tau_{lm}
({\mathbf k'}) \rangle,
\label{vector-source-sym1}
\\
(2\pi)^3 \delta^{(3)} ({\bf k}- {\bf k'}) g(k) &=& -
\frac{i}{4}{\hat k}_q [\epsilon_{jmq}  {\hat k}_i {\hat
k}^\prime_l +\epsilon_{ilq} {\hat k}_j {\hat k}^\prime_m] \langle
\tau^{\star}_{ij}({\mathbf k}) \tau_{lm} ({\mathbf k'})
\rangle. \label{vector-source-hel}
\end{eqnarray}
The functions $f(k)$ and $g(k)$ are evaluated in Appendix A,
\bea
f(k) & \simeq & {\mathcal F}_B (\lambda k_D)^{2n_B +3} \
\left [
1+\frac{n_B}{n_B+3}\left(\frac{k}{k_D}\right)^{2n_B+3}
\right ]-
{\mathcal F}_H (\lambda k_D)^{2n_H +3}
\left [
1+\frac{n_H-1}{n_H+4}\left(\frac{k}{k_D}\right)^{2n_H+3}
\right ],
\label{V-S-Source}
\\
g(k)  & \simeq &
{\mathcal G} \lambda k (\lambda k_D)^{n_B+n_H+2}
\left[ 1 + 
\frac{n_H-1}{n_B+3} 
\left (\frac{k}{k_D} \right )^{n_B+n_H+2}
\right],
\label{V-A-Source}
\eea
for $k < k_D$; $f(k)$ and $g(k)$ vanish for $k>k_D$.  
 Here ${\mathcal F}_{B}$, ${\mathcal F}_H$ and ${\mathcal G}$
are constants that depend
on the magnetic field power spectrum indexes $n_B$ and $n_H$, and the
power spectrum normalization,
\bea
&&
{\mathcal F}_B=\frac{\lambda^3 B_\lambda^4}{16 (2n_B+3)\Gamma^2(n_B/2+3/2)},~~~~~
~~~~{\mathcal F}_H=\frac{\lambda^3 H_\lambda^4}{24(2n_H+3)
\Gamma^2(n_H/2+2)},
\nonumber
\\
&&~~~~~~~~~~~~~ {\mathcal G}=\frac{\lambda^3 B^2_\lambda
H^2_\lambda} {24 (n_B+n_H+2)\Gamma(n_B/2+3/2) \Gamma(n_H/2+2)}.
\label{constants} \eea The contribution of magnetic field
helicity to the symmetric source $f(k)$ is negative (as  in the
case for tensor perturbations \cite{caprini03}). The magnetic
source terms in Eqs.~(\ref{V-S-Source}) and (\ref{V-A-Source})
vanish on scales smaller than the cutoff scale $\lambda
<\lambda_D$ because of magnetic field damping. The
singularities at $n_B = - 3/2$, $n_H = -3/2$ and $n_B+n_H=-2$ in
Eqs.~(\ref{constants}) are removable \cite{mack02,
caprini03}. For $n_B >-3/2$ the terms proportional to
$k_D^{2n_B+3}$ and $k_D^{2n_H+3}$ dominate in the expression for
$f(k)$ in Eq.~(\ref{V-S-Source}), and the symmetric source term
 depends on the cutoff wavenumber $k_D$ but not on
$k$, and so is a white noise  spectrum \cite{durrer03}.

\subsection{Vorticity perturbations}

A cosmological magnetic field contributes,
 via the linearized Einstein equations, to
all three kinds of perturbations, scalar, vector, and tensor modes
(for a recent review see \cite{lewis04b}).
 Here we focus on
the effects a stochastic magnetic field with helicity has on vector
perturbations.\footnote{See Ref.~\cite{pogosian02} for a
study of helical vorticity fields. They did not account for
magnetic field helicity acting as a source in the helical
vorticity field  perturbation equation of motion. }
 The vector
metric perturbation may be described in terms of two
gauge-invariant divergenceless three-dimensional vector fields,
the vector potential $\bf{V}$ (which is a vector perturbation of
the extrinsic curvature), and a vector parameterizing the
transverse peculiar velocity of the plasma, the vorticity ${\bf
\Omega}={\bf v}-{\bf V}$, where ${\mathbf v}$ is the spatial part
of the four-velocity perturbation of a stationary fluid element
\cite{durrer98}. In the absence of a
source a vector perturbation decays with time and so can be
ignored.

Since electromagnetism is conformally invariant it is possible to
rescale fields by appropriate powers of the scale factor and
simply obtain Maxwell's equations in the expanding universe from
the Minkowski spacetime Maxwell equations. Since the fluid
velocity is small the displacement current in Amp\`ere's law
may be neglected; this implies the current $ {\bf J} =\bf{ \nabla\times}
{\bf B }/(4\pi)$. The residual ionization is large enough to
ensure that magnetic field lines are frozen into the plasma, so
the induction law takes the form $({\partial /\partial t){\bf B}}
= {\bf  \nabla} \times( {\bf v }\times {\bf B})$.
As a result the baryon Euler equation for $\bf v$ has a Lorentz
force ${\mathbf L({\mathbf x})}\simeq - \left\{{\mathbf B}({\bf
x})\times\left[\nabla\times {\bf B}({\bf x})
\right]\right\}/(4\pi)$ as a source term. The photons are neutral
so the photon Euler equation does not have a Lorentz force source
term. The Euler equations for photons and baryons are
\cite{hu97,mack02},
\begin{eqnarray}
\dot{{\bf \Omega}}_{\gamma}+\dot{\tau}({\bf v}_{\gamma}-{\bf v}_{b})
&=& 0,
\label{eq:V-momentum-photon}\\
\dot{{\bf \Omega}}_{b}+\frac{\dot{a}}{a}{\bf \Omega}_{b}-
\frac{\dot{\tau}}{R}
({\bf v}_{\gamma}-{\bf v}_{b})
&=& \frac{{\bf L}\!^{(V)}\!({\mathbf k})}{a^4(\rho_b+p_b)},
\label{eq:V-momentum-baryon}
\end{eqnarray}
where an overdot represents a derivative with respect to
conformal time $\eta$, and $ {\mathbf\Omega}_{\gamma}= {\mathbf
v}_{\gamma}-{\mathbf V}$ and ${\mathbf\Omega}_{b} = {\mathbf
v}_{b}-{\mathbf V}$
 are the vorticities of the photon and baryon fluids. Here
 $\dot{\tau}=n_e\sigma_Ta$ is the differential optical depth, with
$n_e$ the free electron density and $\sigma_T$  the Thomson
cross section,
$R\equiv(\rho_b+p_b)/(\rho_\gamma+p_\gamma)\simeq3\rho_b/4\rho_\gamma$
is the momentum density ratio between baryons and photons, and
$L^{(V)}_i$ is the transverse vector (divergenceless) part of
the Lorentz force.

The average Lorentz force
$\langle{\bf L(x)}\rangle =-\langle{\bf B} \times [{\bf \nabla}
\times {\bf B}]\rangle/(4\pi)$ vanishes,
while the r.m.s. Lorentz force
$\langle{\bf L (x) \cdot L(x)}\rangle^{1/2}$ is non-zero and acts as a
source in the  vector perturbation equation.
 The magnetic helicity spectrum $P_H(k)$ contributes to the
 symmetric part of the Lorentz force spectrum,
see the expression for  $f(k)$ in Eq.~(\ref{V-S-Source}).
As a result  magnetic
helicity affects the symmetric vorticity perturbation spectrum
via the Euler equation, and so contributes to  parity-even CMB
fluctuations. The helical part of the Lorentz force spectrum is
completely determined by $g(k)$, Eq.~(\ref{V-A-Source}), and acts as
a source for the helical part  of the vorticity perturbation spectrum.
Solving the Euler equations in the tight-coupling limit when
${\bf v}_{\gamma}\simeq {\bf v}_{b}$, we have
\cite{subramanian98b,mack02},{\footnote{We use the ``helicity''
basis of Sec.~1.1.3 of Ref.~\cite{varshalovich89}
and decompose a vector ${\bf A} =
\sum_{\mu =-1}^1 {\bf e}_{(\mu )} A^{(\mu )}$, where ${\mathbf
e}^{(\pm 1)}$ and ${\mathbf e}^{(0)}$ are the unit basis
vectors.  The unit vector ${\mathbf e}^{(0)}$  is chosen to be in
the direction of wave propagation ${\bf e}^{(0)}={\bf {\hat k}}$ and
$ {\mathbf e}^{(\pm 1)}({\bf k}) =-i ({\mathbf e}_1 \pm  i {\mathbf
e}_2)/\sqrt{2}. $}}
\be
\Omega^{(\pm 1)}(\eta,{\mathbf k})\simeq
\frac{k\Pi^{(\pm 1)}({\mathbf k})\eta}{(1+R)(\rho_{\gamma0}+p_{\gamma0})}.
\label{vorticity1}
\ee
Here $p_{\gamma 0}$ and $\rho_{\gamma 0}$
are the photon pressure and energy density today, and the $\Omega^{(0)}$
 component vanishes
due to the transversality condition. This result can also be
obtained from  the Einstein equation \cite{mack02}.

The expression  in Eq.~(\ref{vorticity1})  is valid on scales $\lambda$
larger than the comoving Silk scale $\lambda_S$.
On smaller scales  the photon
viscosity becomes comparable to the effects of the magnetic field
and so must be accounted for in the Euler equation. On these smaller
scales with $k>k_S$ \cite{subramanian98b},
 \be
\Omega^{(\pm 1)}(\eta,{\mathbf k}) \simeq
\frac{\Pi^{(\pm 1)}(\mathbf k)}{(kL_\gamma/5)(\rho_{\gamma0}+p_{\gamma0})},
\label{vorticity-viscosity}
\ee
where $L_\gamma$ is the photon mean free path length.

We define the CMB anisotropies in terms of the vorticity
perturbation power spectrum,{\footnote{ Both the vorticity
$\mathbf{\Omega}$ and the vector potential $\mathbf{V}$ appear in
the equations for the CMB temperature and polarization
fluctuations. The set of the 
 equations 
governing the vector perturbation dynamics 
 is given  in Eqs.~(35)--(38) of Ref.~\cite{lewis04b}. These show 
that the contribution from ${\bf \Omega}$ to the
CMB anisotropies dominates over that from ${\bf V}$, justifing 
 neglect of the ${\bf V}$ contribution to CMB temperature and polarization 
 anisotropies \cite{durrer98,subramanian98b,mack02}.}}
\be \langle
\Omega_i^*({\mathbf k}) \Omega_j ({\mathbf k'})\rangle = (2\pi)^3
\delta^{(3)}({\mathbf k}-{\mathbf k'}) \left[ P_{ij}
|\Omega|^2(k) + i  {\hat k}_q \epsilon_{ijq} \omega (k)\right].
\label{vorticity-spectrum} \ee Here $|{\Omega}|^2(k)$ is the
symmetric part of the vorticity power spectrum, directly related
to the symmetric magnetic field source spectrum $f(k)=|\Pi|^2(k)$,
while  $\omega(\eta,k)$ is the helical part of the vorticity
spectrum and is determined by the helical magnetic field source term
$g(k)$, \bea \Omega(\eta,k) \simeq  \cases {
{\displaystyle\frac{k \eta} {(1+R)(\rho_{\gamma0}+p_{\gamma0})}
\Pi(k), }& $k<k_S$,\cr {\displaystyle
\frac{1}{(kL_\gamma/5)(\rho_{\gamma0}+p_{\gamma0})} \Pi(k),} &
$k>k_S$,\cr }~~~~ \omega(\eta,k) \simeq \cases{ {\displaystyle
\left[\frac{k \eta} {(1+R)(\rho_{\gamma0}+p_{\gamma0})}\right]^2
g(k),}  & $k<k_S$, \cr {\displaystyle\left
[\frac{1}{(kL_\gamma/5)(\rho_{\gamma0}+p_{\gamma0})}\right]^2
g(k),} & $k>k_S$.\cr }  \label{vorticity-sym-spectrum} \eea Here the
factor $1+R$  for $k<k_S$
reflects the suppression of the vorticity field due to the tight
coupling between baryons and photons, because photons, being
neutral,  are not influenced by the magnetic field Lorentz force
source.

\section{parity-even CMB Fluctuations from  magnetic helicity}

Cosmological magnetic field vector and tensor mode contributions
to large angular scale ($l<100$)
 CMB fluctuations are of the same order of magnitude \cite{mack02},
while small angular scale  ($l>100$) CMB fluctuations are
dominated by the vector mode contribution \cite{caprini03}.
Stochastic non-helical magnetic field effects  on the CMB
temperature and polarization anisotropies are discussed  in
detail in Refs.~\cite{mack02,subramanian98b,lewis04b}. Here we
compute the CMB vector mode fluctuation arising from the helical
part of the magnetic field power spectrum (also see
Ref.~\cite{pogosian02}).

To compute the CMB temperature and polarization anisotropy
 power spectra we use
the total angular momentum representation \cite{hu97}. Given CMB
temperature and polarization anisotropy integral solutions of the Boltzmann
temperature equation, the
CMB fluctuation power spectra are \cite{hu97}, \be C^{{\mathcal
X}{\mathcal X}^\prime}_l ={2 \over \pi} \int dk\,k^2 \sum_{m}
{{\mathcal X}_{l(m)}(\eta_0,k) \over 2l+1} {{{\mathcal
X}_{l(m)}^{\prime}}(\eta_0,k) \over 2l+1},
\label{C-power-spectrum} \ee where $\eta_0$ is the conformal time
now and $\mathcal X$ is either $\Theta$, $E$, or  $B$,
which represent respectively the temperature, $E$-polarization, and
$B$-polarization anisotropies.  For the vector mode the sum includes
only terms with $m=\pm1$. In what follows  we use results from
Ref.~\cite{mack02} for the CMB fluctuation power spectra
$C^{{\mathcal X}{\mathcal X}^\prime}_{(S)l}$ induced by the
symmetric (non-helical) part of magnetic field power spectrum and
proportional to $f_B(k) \sim \int d^3\!p\, P_B(p)P_B(|{\bf k-
p}|)$.

The  complete parity-even CMB fluctuation power spectra
may be expressed as,
 \be
C^{{\mathcal X}{\mathcal X}^\prime}_l=
C^{{\mathcal X}{\mathcal X}^\prime}_{(S)l}-
C^{{\mathcal X}{\mathcal X}^\prime}_{(A)l}, 
\label{decomposition_Cl} \ee where the $C^{{\mathcal X}{\mathcal
X}^\prime}_{(A)l}$ are the power spectra induced by magnetic
helicity, i.e., proportional to $f_H(k)$. The minus sign reflects
the negative contribution of
 magnetic helicity  to the total parity-even CMB
fluctuation  power spectra.{\footnote{ Reference \cite{pogosian02}
ignores the magnetic helicity contribution to the symmetric
vorticity power spectrum. That is, in their computation
they ignore  all  terms proportional to
$ \int d^3\!p\,P_H(p)P_H(|{\bf k- p}|)$.}} This result
also holds for the tensor mode case \cite{caprini03}. The
fractional difference $\kappa^{{\mathcal X}{\mathcal X'}}_l \equiv
1-C_{(A)l}^{{\mathcal X}{\mathcal X}^\prime}/
C_{(S)l}^{{\mathcal X}{\mathcal X}^\prime}$, where
$0<\kappa^{{\mathcal X}{\mathcal X'}}_l<1$, can be used
to characterize the
 reduction of the parity-even CMB fluctuation power
spectra amplitudes as a consequence of non-zero magnetic helicity. The ratio
$C_{(A)l}^{{\mathcal X} {\mathcal X}^\prime}/ C_{(S)l}^{{\mathcal X}
{\mathcal X}^\prime}$ may be expressed  in terms of $g(k)/f(k)$, i.e., in
terms of $P_{0H}/P_{0B}$ and spectral indexes $n_H$ and $n_B$. We
find that for any parity-even CMB fluctuation power spectrum \be
\kappa^{\mathcal {X X^\prime}}_l = 1-\frac{2(2n_B+3)}{3(2n_H+3)}
\left(\frac{P_{H0} k_D^{n_H-n_B}}{P_{B0}}\right)^2{\mathcal
R}^{{\mathcal X}{\mathcal X}}_a (n_B, n_H, l)~. \label{chi} \ee
Here ${\mathcal R}^{\mathcal {X X^\prime}}_a(n_B, n_H, l)$ are
dimensionless functions of $l$ and spectral indexes $n_B$ and
$n_H$,
 and the index  $a=1$ and $a=2$ corresponds to $n_H >-3/2$ and $n_H<-3/2$,
respectively. In this section we present explicit forms for
${\mathcal R}^{{\mathcal X}{\mathcal  X}^\prime}_a.$
Parity-odd CMB fluctuation
power spectra, such as $C_l^{\Theta B}$ and $C_l^{EB}$, receive a
contribution only from the helical part of the magnetic field
source power spectrum, $g(k)$, i.e., from terms proportional to
$\int d^3\!p\,P_B(p)P_H(|{\bf k- p}|)$ and $\int d^3\!p\,P_H(p)
P_B(|{\bf k- p}|)$.

\subsection{CMB temperature anisotropies}

Vector perturbations induce  CMB temperature
anisotropies via  the Doppler and  integrated Sachs-Wolfe effects.
Neglecting a possible dipole contribution from
 the velocity perturbation
$\mathbf{v}$ today,  and using the Boltzmann
 temperature transport equation solutions
$\Theta_l$ for vector perturbations (see Eqs.~(5.2), (5.5), and (5.6)
 of Ref.~ \cite{mack02}), we
get\footnote{For $C^{\Theta\Theta}_{(S)l}$
see Eq.~(5.7) of Ref.~\cite{mack02}.}
\bea C^{\Theta\Theta}_{(A)l} &=&l(l+1)
\frac{(2\pi)^{2n_H+8}~v_{H \lambda}^4}
{6(2n_H+3)\Gamma^2(n_H/2+3/2)}
\frac{(k_D\eta_0)^{2n_H+3}}{(k_\lambda\eta_0)^{2n_H+6}} \nonumber
\\
& & \times
\left[\frac{\eta^2_{\text{dec}}}{(1+R_{\text{dec}})^2}\int^{k_S}_0dk\,k
+\frac{25}{L^2_{\gamma\,\text{dec}}}\int^{k_D}_{k_S}\frac{dk}{k^3}\right]
\left\{1+\frac{n_H-1}{n_H+4}\left(\frac{k}{k_D}\right)^{2n_H+3}\right\}
J^2_{l+1/2}(k\eta_0). \label{V-temp-power-spectrum-1} \eea
Here $J_{l+1/2}(x)$ is a Bessel function, and we have
used the
analogy with Alfv\'en velocity, $v_{A \lambda} \equiv
B_\lambda/\sqrt{4\pi (\rho_{\ga 0} + p_{\ga 0})}$, to introduce
the  ``helicity''
velocity
$v_{H \lambda} \equiv H_\lambda/\sqrt{4\pi (\rho_{\ga 0} + p_{\ga 0})}$.
For $n_H >-3/2$ the integral expressing  the CMB
temperature anisotropy in Eq.~(\ref{V-temp-power-spectrum-1}) is
dominated by the first term ($1$) in the curly brackets,
 while for $n_H<-3/2$ the
second term $\propto (n_H-1)/(n_H+4)$ in the curly brackets  dominates.

The integral in Eq.~(\ref{V-temp-power-spectrum-1}) is  split
into two parts. The first integral is
evaluated using the solution for vorticity that does not  account
for the effects of viscosity, Eq.~(\ref{vorticity1}), while the second
integral makes use of the solution in Eq.~(\ref{vorticity-viscosity}) which
accounts for the effects of viscosity.
However,  numerical integration shows that the first
integral $\int_0^{k_S}$ dominates,
since the second integral with vorticity damped by photon
viscosity (when $k_S<k<k_D$) contributes less than of order
 1 \% for
$l\leq500$, for all values of the spectral indexes $n_{B}$ and
$n_{H}$ \cite{mack02}. Thus the contribution of the second integral may
be safely neglected.
The first integral is evaluated
 using
Eqs.~(5.9)--(5.11) and (5.15) of Ref.~\cite{mack02}, and
Appendix B of this  paper.

Retaining only the leading-order terms  for
$l\leq500$, for $n_H>-3/2$ we find,
\bea {\mathcal R}_{1}^{\Theta
\Theta}\simeq \cases { 1~, & ~~~~~~~~$n_B>-3/2$, \cr {\displaystyle
\frac{2(n_B+3)(n_B+2)}{n_B}
\left(\frac{k_D}{k_S}\right)^{2n_B+3}~, }& $-2<n_B<-3/2$, \cr
{\displaystyle \frac{2^{2n_B+6} (n_B+3)
\Ga(-2n_B-4)}{n_B\Gamma^2(-n_B-2)}\left(\frac{k_S
\eta_0}{l}\right) \left(\frac{k_D\eta_0}{l}\right)^{2n_B+3}~, }&
$-3<n_B<-2$. \cr } \label{4.110} \eea Even for a magnetic field with
maximal
helicity (when  $|P_H(k)|=P_B(k)$), for $n_B \simeq
n_H >-3/2$~ the ratio ${C_{(A)l}^{\Theta \Theta}}/{C_{(S)l}^{\Theta
\Theta}}=2/3$ ($\kappa^{\Theta \Theta}_l =1/3$) and the CMB temperature
anisotropy power spectrum $C_l^{\Theta \Theta}>0$.
 For $n_B>-2$ the function ${\mathcal R}_{1}^{\Theta
\Theta}(n_B, n_H, l)$ is independent of $l$.\footnote{For
$n_B=-2$ there is a weak dependence on $l$, ${\mathcal
R}_1^{\Theta \Theta} (n_B, n_H, l) \sim 1/\mbox{ln}(k_S\eta_0
/l)$.} For ~$-3 < n_B < -2$, ${\mathcal R}_{1}^{\Theta \Theta}
(n_B, n_H, l)$ is a growing function of $l$, scaling as
 $l^{-2n_B-4}$. The maximum growth rate of $l^2$ occurs for $n_B =  -3$.

For $n_H< -3/2$ we distinguish two different regions,
$-2<n_H<-3/2$ and $-3< n_H<-2$. Using
Eqs.~(\ref{eq:GR-6.574.2})--(\ref{bessel-int3}) (also see
Eqs.~(5.10)--(5.12) and (5.15) of Ref.~\cite{mack02}), and
$k_S \eta_0 \gg l$, we find,
\bea {\mathcal R}^{\Theta \Theta}_{2} \simeq \cases {
{\displaystyle \frac{(n_H-1)(n_B+3)(n_B+2)}{n_B(n_H+4)(n_H+2)}
\left(\frac{k_D}{k_S}\right)^{2(n_B-n_H)}\!\!, } 
~~~~~~~~~~~~~~~~~~~~~~~~~~~~~~~~~~~~~~~~~~~~~~~~ -2<n_B<-3/2, \cr
{\displaystyle
 \frac{2^{2n_B+6}(n_H-1)(n_B+3)\Gamma(-2n_B-4)}
{n_B(n_H+4)(n_H+2)\Gamma^2(-n_B-2)}
\left(\frac{k_D}{k_S}\right)^{2(n_B-n_H)}\left(\frac{k_S\eta_0}{l}
\right)^{2n_B+4}\!\!, }   ~~~~n_B<-2< n_H <-3/2 , \cr {\displaystyle
\frac{2^{2n_B-2n_H-1}(n_H-1)(n_B+3)\Gamma(-2n_B-4)\Gamma^2(-n_H-2)}
{n_B(n_H+4) \Gamma(-2n_H-4)\Gamma^2(-n_B-2)} \left(\frac{k_D
\eta_0}{l}\right)^{2(n_B-n_H)}\!\!, } ~~~~~~~~n_B \leq  n_H <-2. \cr }
%
\label{4.12} \eea The function
${\mathcal R}_{2}^{\Theta \Theta}$ is independent of $l$
for ~$-2<n_B \leq n_H
<-3/2$, while  ${\mathcal R}_{2}^{\Theta \Theta}$ grows
 as $ l^{-2n_B-4}$ for ~$-3
<n_B < -2 \leq  n_H <-3/2$, which coincides with the growth rate of
 ${\mathcal R}_{1}^{\Theta \Theta} (n_B, n_H, l)$ for
$-3<n_B<-2$ and $n_H > -3/2 $. Thus if $n_B<-2$, independent
of whether $n_H >
-3/2$ or ~$-2 <n_H < -3/2$, ${\mathcal R}_a^{\Theta \Theta} (n_B,
n_H, l) \propto l^{-2n_B-4}$. For ~$-3< n_B \leq n_H <-2$ the
function ${\mathcal R}_{2}^{\Theta \Theta} (n_B, n_H, l)$
monotonically increases with $l$  if  $n_H >n_B$, while it is
independent of $l$ for $n_H=n_B$.

The approximate expressions  above are accurate to better than
15 \% for $n_H
> -2$, to better than 30 \% for ~$-2.5\leq n_H\alt-2$, and to better than
a few percent for
$n_H<-2.5$
 \cite{mack02}. We do not reproduce the explicit forms
of the power spectra for the cases $n_H=-3/2$ or $n_H=-2$, and
$n_B=-3/2$ or $n_B=-2$, since the corresponding results can be
easily derived via a straightforward extension of the computation
presented in Ref.~\cite{mack02}; see Eq.~(5.16) there.

\subsection{CMB polarization anisotropies}

Vector perturbations generate both  $E$ and $B$  CMB
polarization  anisotropies \cite{kamionkowski97b,review4}. Since
scalar perturbations do not induce  a magnetic ($B$) CMB
polarization anisotropy, the future detection of a $B$ polarization signal
will indicate the presence of a vector and/or a tensor
perturbation mode. In this subsection  we consider the
CMB polarization anisotropies that result
from vector perturbations induced by an helical
cosmological magnetic field.

\subsubsection{$E$-polarization}

Using the integral solution  for the CMB electric
($E$) polarization $E_l$, see Eq.~(6.6)  of
Ref.~\cite{mack02}, the CMB $E$-polarization power spectrum from
magnetic helicity is,\footnote{Just as for the temperature
integral solution $\Theta_l$,  $E_l$  is  also
expressed in terms of $\mathbf \Omega$. To compute $C_l^{EE}$ we
use Eqs.~(\ref{C-power-spectrum}) and
(\ref{vorticity-sym-spectrum}) along with Eq.~(\ref{V-S-Source}). An
 expression for $C_{(S)l}^{EE}$ is given in Eqs.~(6.7) and (6.9)--(6.12) of
Ref.~\cite{mack02}.}
\bea
C^{EE}_{(A)l}
&=&(l-1)(l+2)\frac{(2\pi)^{2n_H+8}~v_{H \lambda}^4}{54(2n_H+3)
\Gamma^2(n_H/2+2)}
\frac{(k_D\eta_0)^{2n_H+3}}{(k_\lambda\eta_0)^{2n_H+6}}
L^2_{\gamma\,\text{dec}}
\left(\frac{\eta_{\text{dec}}\eta_0}{1+R_{\text{dec}}}\right)^2
\nonumber\\&&
\times\int^{k_S}_0dk\,k^5
\left[1+\frac{n_H-1}{n_H+4}\left(\frac{k}{k_D}\right)^{2n_H+3}\right]
\left[(l+1)\frac{J_{l+1/2}(k\eta_0)}{(k\eta_0)^2}
-\frac{J_{l+3/2}(k\eta_0)}{k\eta_0}\right]^2.\label{V-E-power-spectrum}
\end{eqnarray}
The contribution from  the integral  $\int_{k_S}^{k_D}$ (i.e.,
 the contribution from the
region where vorticity is damped by photons) is negligible.
When evaluating the integral in Eq.~(\ref{V-E-power-spectrum})
 we retain only the
dominant term, and this differs depending
 on whether $n_H>-3/2$ or $n_H<-3/2$.
 We also use Eq.~(\ref{jjj}) to approximate the cross term
$J_{l+1/2}(k\eta_0)J_{l+3/2}(k\eta_0)$.
To evaluate  ${\mathcal R}^{EE}_a(n_B, n_H, l)$ we consider the
following three regions:
(i) $n_B>-3/2$ and $n_H>-3/2$; (ii) $-3<n_B<-3/2$ and $n_H >-3/2$;
and, (iii) $-3 < n_B <-3/2$ and $n_H<-3/2$.

For $n_H>-3/2$, using
Eqs.~(\ref{bessel-int})--(\ref{bessel-int2}) and
Eq.~(\ref{bessel-int3}) and retaining only
 the leading terms in the limit where  $k_S
\eta_0 \gg l$, we find,  \bea {\mathcal R}^{EE}_{1} \simeq \cases {1~, &
~~~~~~~~$n_B
>-3/2$ \cr {\displaystyle \frac{2(n_B+3)^2}{n_B}\left(
\frac{k_D}{k_S}\right)^{2n_B+3},} & $-3<n_B<-3/2$. \cr}
\label{4.20} \eea The function ${\mathcal
R}^{EE}_{1}={\mathcal R}^{\Theta \Theta}_{1}$ for $n_B>-3/2$, and
${\mathcal R}^{EE}_{1}=(n_B+3){\mathcal R}^{\Theta \Theta}_{1}$
for $-2<n_B<-3/2$. The expression for ${\mathcal R}^{EE}_{1}$ is
the same for all values of $n_B$ in the range  $-3<n_B<-3/2$, while
 ${\mathcal R}^{\Theta \Theta}_{1}$ differs
 depending on whether  $-2<n_B<-3/2$ or $-3<n_B<-2$, see Eq.~(\ref{4.110}).

For $n_H<-3/2$  we need to consider the ranges $-2<n_H<-3/2$ and
$n_H<-2$ separately \cite{mack02}. Using
Eqs.~(\ref{bessel-int})--(\ref{bessel-int2}) and
Eq.~(\ref{bessel-int3}), we find that for $n_H<-2$ the expression
for $C^{EE}_{(A)l}$ is identical to the one in the region
$-2<n_H<-3/2$ to within 15 \% accuracy. This simplifies the
computation,  and for $-3<n_B\leq n_H<-3/2$ we find, \be
 {\mathcal R}^{EE}_{2} (n_B, n_H, l) \simeq
\frac{(n_H-1)(n_B+3)^2}{n_B(n_H+4)(n_H+3)}
\left(\frac{k_S}{k_D}\right)^{2(n_H-n_B)} , \label{RE2} \ee while for
$-2<n_B \leq n_H <-3/2$ we have $ {\mathcal R}^{EE}_{2} (n_B,
n_H, l)= {(n_B+3)(n_H+2)}/({(n_B+2)(n_H+3)}) {\mathcal R}^{\Theta
\Theta}_{2} (n_B, n_H, l)$, where $ {\mathcal R}^{\Theta
\Theta}_{2}$ is given in Eq.~(\ref{4.12}).

\subsubsection{$B$-polarization}
To compute the contribution from cosmological magnetic helicity to the CMB
 $B$-polarization power spectrum we use the integral
solution for $B_l$,  Eq.~(6.16) of Ref.~\cite{mack02}. The
CMB $B$-polarization
power spectrum is,
\begin{eqnarray}
C_{(A)l}^{BB} &=& (l-1)(l+2) \frac{(2\pi)^{2n_H+8}~ v_{H
\lambda}^4}{54(2n_H+3)\Gamma^2(n_H/2+2)}
\frac{(k_D\eta_0)^{2n_H+3}}{(k_\lambda\eta_0)^{2n_H+6}}
L^2_{\gamma\,\text{dec}}
\left(\frac{\eta_{\text{dec}}\eta_0}{1+R_{\text{dec}}}\right)^2
\nonumber\\&&
\times\int^{k_S}_0dk \, k^5
\left[1+\frac{n_H-1}{n_H+4}\left(\frac{k}{k_D}\right)^{2n_H+3}\right]
\frac{J^2_{l+1/2}(k\eta_0)}{(k\eta_0)^2}.
\label{V-B-power-spectrum}
\end{eqnarray}
 Here again the contribution from the
 region where viscous effects are important
is negligibly small.
 An expression for $C_{(S)l}^{BB}$ is given in  Eqs.~(6.18)--(6.19)
of Ref.~\cite{mack02}.

Using Eq.~(\ref{bessel-int}) and retaining the leading  terms
  for $n_H<-3/2$,
we get, \be {\mathcal R}^{BB}_{1} \simeq
{\mathcal R}^{EE}_{1}, ~~~~~~~~~~~~~~~~~~~~~~~~~ {\mathcal
R}^{BB}_{2} \simeq {\mathcal R}^{EE}_{2}, \label{4.20B} \ee
to  better than 20 \% accuracy \cite{mack02}.

Both functions $R^{EE}_a$ and $R^{BB}_a$ (i.e., the coefficients
$\kappa^{EE}_l$ and $\kappa^{BB}_l$) are independent of $l$. This
means that cosmological magnetic helicity reduces the CMB polarization power
spectrum amplitudes by the same scale factor for the $E$-
and the $B$-polarizations. The ratio between contributions to the $E$-
or $B$-polarization
signal from the helical and the symmetric parts of the magnetic field
are independent of $l$ for the entire range of spectral
indexes in the case of the vector mode, while for the tensor mode case
 the ratios depend on $l$ for $n_H<-2$ \cite{caprini03}.

\subsubsection{Temperature-$E$-polarization cross correlation}

We may obtain the CMB temperature--$E$-polarization cross-correlation
 power spectrum
$C_l^{\Theta E}$ from the integral solutions for the
 temperature  and $E$-polarization anisotropies. 
Like $C_l^{\Theta \Theta}$ this power spectrum
  is also parity-even and  only the
symmetric part of the magnetic field source $f_B(k)$
(which also contains a contribution from the helical part of the
magnetic field power spectrum  $P_H(k)$) contributes to it. As
 discussed in Ref.~\cite{hu97} (see Fig.~5 there),
 the vector dipole temperature anisotropy radial function
$j^{(1V)}_l=j_l\sqrt{l(l+1)/2}/x $ does not correlate well with
its $E$-polarization anisotropy radial function
$\epsilon^{(V)}_l=[j_l/x^2 + j_l^\prime/x] \sqrt{(l-1)(l+2)}/2$
(here $j_l^\prime(x)$ is the partial derivative with respect to $x$
 of the Bessel function of
fractional order $j_l$),  while
the vector quadrupole temperature anisotropy radial function
$j^{(2V)}_l=\sqrt{3l(l+1)/2}(j_l/x)^\prime$ does. To compute
the vector mode CMB temperature--$E$-polarization
cross-correlation power spectrum we therefore have to
 retain the term proportional to
$j^{(2V)}_l$ in the vector
temperature integral solution $\Theta_l$
which was neglected previously\footnote{Since it is suppressed
relative to the term proportional to $j^{(1V)}_l$.} in the derivation of
$C_l^{\Theta \Theta}$ (see
Eq.~(\ref{V-temp-power-spectrum-1})). With the $j^{(2V)}_l$ term
 the CMB
temperature anisotropy integral solution  is \cite{hu97},
\begin{equation}
\frac{\Theta^{(V)}_l(\eta_0,k)}{2l+1}\simeq
\sqrt{\frac{l(l+1)}{2}}\Omega(\eta_{\text{dec}},k)
\left[
\frac{j_l(k\eta_0)}{k\eta_0}+
\frac{kL_{\gamma\,\text{dec}}}{3}
\left\{
(l-1)\frac{j_l(k\eta_0)}{(k\eta_0)^2}
-\frac{j_{l+1}(k\eta_0)}{k\eta_0}\right\}
\right].
\label{V-temp-int-soln-4}
\end{equation}
$C_{(S)l}^{\Theta E}$ is given in Eq.~(7.3) of
Ref.~\cite{mack02}, and for $C_{(A)l}^{\Theta E}$ we get,
\begin{eqnarray}
C^{\Theta E}_{(A)l} =&-&\sqrt{l(l-1)(l+1)(l+2)}
\frac{(2\pi)^{2n_H+8}~v_{H \lambda}^4} {18
(2n_H+3)\Gamma^2(n_H/2+2)}
\frac{(k_D\eta_0)^{2n_H+3}}{(k_\lambda\eta_0)^{2n_H+6}}
\left(\frac{\eta_{\text{dec}}\eta_0}{1+R_{\text{dec}}}\right)^2L_{\gamma\,\text{dec}}
 \nonumber\\ &\times&
\int^{k_S}_0dk\,k^4
\left[
1+\frac{n_H-1}{n_H+4}\left(\frac{k}{k_D}\right)^{2n_H+3}
\right]
\left[
(l+1)\frac{J^2_{l+1/2}(k\eta_0)}{(k\eta_0)^3}
-\frac{J_{l+1/2}(k\eta_0)J_{l+3/2}(k\eta_0)}{(k\eta_0)^2}\right.+
\nonumber\\
& &\left.\mbox{}~~~~~~~~~~~~~~~+\frac{kL_{\gamma\,\text{dec}}}{3}
\left\{
(l^2-1)\frac{J^2_{l+1/2}(k\eta_0)}{(k\eta_0)^4}
-2l\frac{J_{l+1/2}(k\eta_0)J_{l+3/2}(k\eta_0)}{(k\eta_0)^3}
+\frac{J^2_{l+3/2}(k\eta_0)}{(k\eta_0)^2}\right\}\right].
\label{V-cross-power-spectrum}
\end{eqnarray}
The first two terms in the second pair of square brackets in this integral
  arise from the correlation of $j^{(1V)}_l$ with $\epsilon^{(V)}_l$.
A numerical evaluation  of the integral
 \cite{mack02} shows that these two terms
roughly cancel each other as a consequence of the low
 correlation between $j^{(1V)}_l$
and $\epsilon^{(V)}_l$ \cite{hu97}.
 This may also be seen by using
Eqs.~(\ref{bessel-approx}) and (\ref{bessel-approx1}),
\be \frac{l+1}{x}J_{l+1/2}^2(x) - J_{l+1/2}(x)J_{l+3/2}(x)
\simeq \frac{1}{2}[J_{l-1/2}(x)J_{l+1/2}(x)-
J_{l+1/2}(x)J_{l+3/2}(x)]\simeq \sin(2x-l\pi)- \sin(2x-l\pi-\pi)
\simeq 0. \label{TEap} \ee In what follows we neglect
these two terms.  We use
Eqs.~(\ref{bessel-int})--(\ref{bessel-int2}) to evaluate the three terms in the curly
brackets of Eq.~(\ref{V-cross-power-spectrum}). The terms from the
correlation between $j^{(2V)}_l$ and $\epsilon^{(V)}_l$ are
suppressed by an additional factor of $kL_{\gamma\,\text{dec}}$,
relative to the two terms from the correlation between
$j^{(1V)}_l$ and $\epsilon^{(V)}_l$. In the limit $l\gg1$, these
terms and the squared sum of the two Bessel function terms
 in the expression for
 $C_l^{EE}$ in
Eq.~(\ref{V-E-power-spectrum})
 (the last factor inside the integral of this equation)
are almost identical. Thus apart
from an overall minus sign, the $C_l^{\Theta E}$
are approximately equal to the corresponding $C_l^{EE}$ \cite{mack02}.
Here  our approximation might not be as accurate because, accounting for the
suppression factor $kL_{\gamma\,\text{dec}}$, 
 the neglected contribution from the
correlation  between $j^{(1V)}_l$ and $\epsilon^{(V)}_l$ 
could be comparable to the retained
 contribution from
the correlation between $j^{(2V)}_l$ and $\epsilon^{(V)}_l$.

\section{Parity-Odd CMB Fluctuations from  Magnetic Helicity}

Magnetic helicity induces  parity-odd cross correlations
between the $E$- and  $B$-polarization anisotropies,
 as well as between temperature
and $B$-polarization anisotropies \cite{pogosian02,caprini03}. Such
off-diagonal parity-odd cross correlations also occur in the
case of an homogeneous magnetic field from the  Faraday rotation
effect \cite{scoccola04}, but not in the case of a
stochastic magnetic field, even one with non-zero helicity \cite{far}.
 Faraday rotation measurements cannot be used to detect
magnetic helicity \cite{ensslin03,campanelli04,far}.
 A possible way of detecting
magnetic helicity directly from CMB fluctuation data
 is  to detect the  above
parity-odd CMB correlations or  to detect the effects magnetic
 helicity has on
 parity-even CMB fluctuations.
 In this Section we study the parity-odd CMB cross-correlations generated
 from vorticity perturbations. The corresponding tensor mode contributions
  are derived
in Sec. VI of Ref.~\cite{caprini03}.

\subsection{Temperature--$B$-polarization cross correlation}

To compute the cross correlation between the CMB temperature and
 $E$-polarization anisotropies we
use the integral solutions for $\Theta_l$ and $B_l$ given
in Eqs.~(5.6) and (6.16) of Ref.~\cite{mack02} and  find,
  \bea C_l^{\Theta
B}&=&- \sqrt{(l-1)l(l+1)(l+2)} \frac{(2\pi)^{n_B+n_H+8}~v_{A
\lambda}^2 ~v_{H \lambda}^2}{18(n_B+n_H+2)\Gamma(n_B/2+3/2)
\Gamma(n_H/2+2)} \frac{(k_D\eta_0)^{n_B+n_H+2}}
{(k_\lambda\eta_0)^{n_B+n_H+6}} \nonumber
\\
&&\times \left[\frac{\eta_{\rm{dec}} \eta_0}{1+R_{dec}}\right]^2
\frac{L_{\gamma \rm{dec}}}{\eta_0}
\int_0^{k_S}dk ~k^3 \left[
1 +
\frac{n_H-1}{n_B+3} 
\left (\frac{k}{k_D} \right )^{n_B+n_H+2}
\right]
J_{l+1/2}^2(k\eta_0)~. \label{T-B-int} \eea
To evaluate this integral it is helpful
 to consider separately the index ranges
 (i)  $n_B+n_H>-2$ and  (ii) $n_H+n_B<-2$.

When $n_B+n_H>-2$  the integral on
the r.h.s. of
Eq.~(\ref{T-B-int}) is dominated by the first term in the square  brackets.
 Using Eq.~(\ref{bessel-int})
with $p=3$ we find, \bea C_l^{\Theta B}=-l^2 \frac{(2\pi)^{n_B+n_H+7}~
v_{A\lambda}^2~v_{H \lambda}^2}{27(n_B+n_H+2) \Gamma(n_B/2+3/2)
\Gamma(n_H/2+2)}
\frac{(k_D\eta_0)^{n_B+n_H+2}}{(k_\lambda\eta_0)^{n_B+n_H+6}}
 \left[\frac{\eta_{\rm{dec}}}{(1+R_{\rm{dec}})\eta_0}\right]^2
\frac{L_{\gamma \rm{dec}}}{\eta_0} (k_S\eta_0)^3. \label{TB-sol1} \eea

When $n_H+n_B<-2$   the integral on the r.h.s.
of
Eq.~(\ref{T-B-int}) is dominated by the second term ($\propto
k^{n_B+n_H+2}$) in the square brackets.
For $n_B+n_H<-5$ the integral in Eq.~(\ref{T-B-int}) ($\int_0^{k_S}
dk k^{n_B+n_H+5}J^2_{l+1/2}(k\eta_0)$) converges for large $k_S$ which can
then be extended to $\infty$, and so the integral
 may be evaluated using  Eq.~(\ref{eq:GR-6.574.2}). We find for
$-6<n_B+n_H<-5$,
\bea
 C_l^{\Theta B}&=& -l^{n_B+n_H+7}
\frac{(2\pi)^{n_B+n_H+8}~2^{n_B+n_H+4}~v_{A\lambda}^2~v_{H
\lambda}^2}{9(n_B+n_H+2) \Gamma(n_B/2+3/2) \Gamma(n_H/2+2)}
\frac{1} {(k_\lambda\eta_0)^{n_B+n_H+6}}
\left[\frac{\eta_{\rm{dec}}}{(1+R_{\rm{dec}})\eta_0}\right]^2
\frac{L_{\gamma \rm{dec}}}{\eta_0}  \nonumber\\ && \times 
\left(\frac{n_H-1}{n_B+3}\right) 
\frac{\Gamma(-n_B-n_H-5)}{\Gamma^2(-n_B/2-n_H/2-2)}.
\label{TB-sol3}
\eea
For $n_B+n_H>-5$ (but still $n_B+n_H<-2$)
the integral in Eq.~(\ref{T-B-int}) diverges at large $k_S$  and so
 the upper
limit cannot be replaced by $\infty$, and  the integral
  cannot be evaluated by using  Eq.~(\ref{eq:GR-6.574.2}). Instead
we approximate it by using Eq.~(\ref{bessel-int}). We find for
$ -5<n_B+n_H<-2$,
\bea  C_l^{\Theta B}&=&
-l^2\frac{(2\pi)^{n_B+n_H+7} ~v_{A\lambda}^2~v_{H
\lambda}^2}{9(n_B+n_H+2)(n_B+n_H+5)\Gamma(n_B/2+3/2)
\Gamma(n_H/2+2)}  \frac{(k_S\eta_0)^{n_B+n_H+5}}
{(k_\lambda\eta_0)^{n_B+n_H+6}} 
\left[\frac{\eta_{\rm{dec}}}{(1+R_{\rm{dec}})\eta_0}\right]^2
\frac{L_{\gamma \rm{dec}}}{\eta_0} \nonumber\\&&\times
\left(\frac{n_H-1}{n_B+3} \right).
\label{TB-sol2} \eea
When $n_B+n_H=-5$ the integration can be done by using
 Eq.~(\ref{bessel-int}) with $p=0$.

At large angular scales ($l<100$) where the contribution from
the tensor mode is significant, for $n_B+n_H>-2$ the  vector mode
${C_l^{\Theta B (V)}}$ and the tensor mode
${C_l^{\Theta B (T)}}$ (see Eq.~(98) of Ref.~\cite{caprini03})
have the same $l$ dependence $\propto l^2$.
For all other values of
spectral indexes $n_{B}$ and $n_{H}$, the growth rate (with $l$) of
 ${C_l^{\Theta B (V)}}$  is faster than
${C_l^{\Theta B (T)}}$.
 In particular,
when the integral in Eq.~(\ref{T-B-int})
 converges at large $k_S$ (for $-6<n_B+n_H<-5$),
${C_l^{\Theta B (V)}}/{C_l^{\Theta B (T)}} \propto l^3$.
When  ~$-5<n_B+n_H<-2$, the integral for the tensor mode
temperature--$B$-polarization cross-correlation power spectrum converges
at large $k$ (see Eq.~(97) of Ref.~\cite{caprini03}),
while it diverges for vorticity perturbations,
Eq.~(\ref{T-B-int}), resulting in
 ${C_l^{\Theta B (V)}}/{C_l^{\Theta B (T)}} \propto
l^{-n_B-n_H-2}$.
 The ratio between  temperature--$B$-polarization
signals from vector and tensor modes is independent
of the amplitudes
of the average magnetic field ($B_\lambda$) and
average magnetic helicity ($H_\lambda$).

For maximally helical magnetic fields with $n_H \simeq n_B$, due 
to the suppression factor $L_{\gamma,{\rm{dec}}}/\eta_0$ 
the temperature-$E$-polarization cross-correlation power spectrum 
$C_l^{\Theta E}$ is smaller the  temperature-$B$-polarization 
cross-correlation power spectrum 
$C_l^{\Theta B}$,\footnote{If $n_H \simeq n_B >-3/2$, 
$C_l^{\Theta E}/C_l^{\Theta B} \simeq  L_{\gamma,{\rm{dec}}}/(2\eta_0)$}
 $C_l^{\Theta E}  \ll 
C_l^{\Theta B }$, but both are $ \propto l^2$, if $n_B+n_H>-5$. The same 
suppression factor makes $C_l^{\Theta B}$   smaller than 
$C_l^{\Theta \Theta}$. For an arbitrary helical field  
 $C_l^{\Theta B}/C_l^{\Theta E}$ depends on the ratio $(P_{H0}/
P_{B0})k_D^{n_H-n_B}$ and order unity  prefactors that depend on 
$n_B$ and  $n_H$.
 A dependence on $l$ appears 
only if $n_B+n_H<-5$ when  the ratio 
$C_l^{\Theta B}/C_l^{\Theta E}$ decreases as  $\propto l^{n_B+n_H+5}$.

\subsection{$E$- and $B$-polarization cross correlation}
To compute cross correlations between  $E$- and $B$-polarization
anisotropies
we use
 the integral solutions for $E_l$ and $B_l$ given in
  Eqs.~(6.14) and (6.16) of
Ref.~\cite{mack02}. We find,
\bea C_l^{E
B}=-(l-1)(l+2)\frac{(2\pi)^{n_B+n_H+8}v_{H \lambda }^2 v_{A \lambda}^2
\eta_0^5}{54 (n_B+n_H+2)
\Gamma(n_B/2+3/2) \Gamma(n_H/2+2)} \frac{(k_D\eta_0)^{n_B+n_H+2}}
{(k_\lambda\eta_0)^{n_B+n_H+6}}
\left[\frac{\eta_{\rm{dec}} \eta_0}{1+R_{\rm{dec}}}\right]^2
L_{\gamma \rm{dec}}^2 
\nonumber
\\
 \times  \int_0^{k_S}dk ~k^5 \left [ 1 + 
\frac{n_H-1}{n_B+3} \left
(\frac{k}{k_D} \right )^{n_B+n_H+2} \right]
\left[(l+1)\frac{J_{l+1/2}^2(k\eta_0)}{(k\eta_0)^3}-
\frac{J_{l+1/2}(k\eta_0)J_{l+3/2}(k\eta_0)}{(k\eta_0)^2}
\right]. \label{E-B-int} \eea The combination of
 Bessel functions
 in the second set of  square brackets in this integral is identical
  to the first two terms  in the second set of
 square brackets in Eq.~(\ref{V-cross-power-spectrum}),
which is negligibly small according to Eq.~(\ref{TEap}). Also the
$E$- and $B$-polarization anisotropy cross-correlation power spectrum
has an additional  suppression factor of
$kL_{\gamma, \rm{dec}}$ relative to the expression in
 Eq.~(\ref{V-cross-power-spectrum}).
 This implies
$C_l^{EB} \ll C_l^{\Theta B}$. Note that this is consistent   with the
result of Ref.~\cite{hu97} that $j^{(1V)}_l$ does not correlate well with
$\epsilon^{(V)}_l.$
The corresponding $C_l^{EB}$ amplitudes in the
tensor mode case (for $l <100$)
are not suppressed and  are of the same order of magnitude as the tensor
mode
 temperature--$B$-polarization anisotropy cross-correlation
 power spectrum \cite{caprini03}.

\section{Conclusion}

In this paper we consider how cosmological magnetic helicity affects
 CMB fluctuations.
 Even for a cosmological magnetic field with maximal helicity, 
 such effects may be detectable  only  if the current magnetic
 field amplitude is at least
$10^{-10}$ or $10^{-9}$ G on Mpc scales. Our analytical expressions for CMB
fluctuation power spectra are valid (to the accuracy of our approximations)
for $n_B>-3$.\footnote{However, a cosmological magnetic field
with spectral index $n_B \simeq 2$
(as might be generated by an MHD cascade in the
early universe \cite{jedamzik04}), has  significant power on small
(galaxy cluster) scales and so measurements of Faraday rotation in clusters
imply an upper limit on the smoothed amplitude $B_\lambda < 10^{-12}$ G on
Mpc scales \cite{dolag,jedamzik04}. Such a ``blue'' cosmological magnetic
field cannot significantly affect CMB anisotropies. A strong 
constraint on magnetic field amplitude on Mpc scales for $n_B>-2$ 
arises from gravitational waves production via a magnetic source, 
if the magnetic field is generated with a power law spectrum 
$P_B(k) = (k/k_{\rm{max}})^{n_B} P_B(k_{\rm{max}})$ 
 (where $k_{\rm{max}} = \eta_{\rm{gen}}^{-1}$ and $\eta_{\rm{gen}}$ 
 is the moment of magnetic field generation)  that can be 
 extrapolated, without damping,  down to the   
 Hubble radius when the magnetic field is generated,  
e.g.,   $\approx  10^{-4}$ Mpc if the magnetic field is generated at the 
 elecroweak phase transition (or even smaller for a magnetic field 
generated during inflation) \cite{caprini00}.}

A cosmological magnetic field generates a $B$-polarization
signal via induced vector and/or tensor
modes, so a  detection of such a signal may indicate the presence
of a  cosmological magnetic field.\footnote{See Refs.~\cite{review4} for 
CMB polarization anisotropy measurements.} However, it has to
be emphasized that a $B$-polarization anisotropy signal can also
  arise in other ways, such as from primordial tensor perturbations
\cite{kamionkowski97b}, gravitational lensing
\cite{lensing}, or Faraday rotation of the CMB anisotropy
polarization plane \cite{kosowsky96,campanelli04,far}.
 The $B$-polarization anisotropy power spectrum
$l^2 C_l^{BB}$ peak position may help identify the
$B$-polarization source. For example, cosmological-magnetic-field-induced
 tensor perturbations only contribute on large
angular scales $l<100$, while $B$-polarization anisotropy from gravitational
lensing has a peak  amplitude   $l^2 C_l^{BB} \sim
10^{-14}$ at $l\sim 1000$ \cite{lensing}. The Faraday rotation
$B$-polarization anisotropy signal from a field with
$B_\lambda =10^{-9}$ G (at $\lambda=1$ Mpc)
 and spectral index $n_B=-2$ peaks at a substantially smaller
 scale $l \sim 10^4$
with a frequency-dependent peak amplitude  $l^2 C_l^{BB} \sim 10^{-12}$ (at
$10$ GHz) and  $l^2 C_l^{BB} \sim 10^{-14}$ (at $30$ GHz) \cite{far}. A
non-helical cosmological magnetic field with
$B_\lambda = 10^{-9}$ G at $\lambda=1$ Mpc   induces
a $B$-polarization anisotropy signal via the vector
perturbation mode with a peak amplitude $l^2 C_l^{BB} \sim 10^{-13}$
  at $l\sim 1000$ \cite{lewis04b}.
 We have shown that a
magnetic field with maximal helicity results in the reduction
 of the $B$-polarization anisotropy
 signal on all scales by a factor of $1/3$ for  $-3/2 < n_B \simeq n_H$,
relative to the  non-helical
magnetic field case.

We have presented analytical expressions for all CMB fluctuation power
spectra affected by cosmological magnetic helicity. Our results show
that cosmological magnetic helicity can affect CMB anisotropies,
 in addition to the  effects
 it has on MHD dynamo amplification and processes in the early universe
\cite{cornwall97,jedamzik04}. It would  be useful to incorporate our
analytical expressions for  $l^2C_l^{{\mathcal X}{\mathcal X'}}$ into
a numerical code (e.g., that of Lewis \cite{lewis04b})  to compute
CMB temperature and polarization anisotropies generated by a
 general cosmological magnetic source.

To set an observational limit on
cosmological  magnetic helicity $H_\lambda$
(or $P_H(k)$) one may proceed as follows. The first step
is to determine the average
magnetic field $B_\lambda$ (or
 $P_B(k)$) using  measurements of the
Faraday rotation of the CMB polarization plane
\cite{kosowsky96,campanelli04,far}. Then one may use
 measurements of the parity-odd temperature-$B$-polarization anisotropies
cross-correlation power spectrum $C_{l}^{\Theta B}$ for $l>100$
(to insure that the tensor mode does not contribute).\footnote{
 Magnetic helicity is the only known source of 
diagonal temperature-$B$-polarization anisotropies
cross correlations. A homogeneous magnetic field
 induces off-diagonal $\Theta - B$ cross correlations
\cite{scoccola04}, but not diagonal correlations.
Faraday rotation does not induce  diagonal or off-diagonal
$\Theta - B$ cross-correlations \cite{far}.}
 According to Eq.~(\ref{T-B-int}),
 if $B_\lambda$ and $n_B$ are known, $C_l^{\Theta B}$ is
 determined by  $H_\lambda$
and $n_H$ (i.e., the helical part of the magnetic field spectrum).
A future detection of $\Theta - B$  cross correlations
 may be used to constrain
$H_\lambda$.  It should be emphasized that on scales $l>100$ magnetic-field-induced
cross correlations between $E$- and $B$-polarization are negligibly small,
 $C_l^{E B}\ll C_l^{\Theta B}$, which can be used as a cross-check of the
source of $B$-polarization anisotropy.
To bound  the range of $n_H$ for a given $n_B$,
 the $l$ dependence of the ratio
$C^{\Theta \Theta}_{(A)l}/C^{\Theta \Theta}_{(S)l}$ can be used.
 In particular, if,  for  $-3<n_B<-2$, $C^{\Theta
\Theta}_{(A)l}/C^{\Theta \Theta}_{(S)l}$ is an increasing function
of $l$ growing as $l^{-2n_B-4}$, then $n_H>-
2$, while if it scales as $l^{2(n_H-n_B)}$ then  $n_B \leq
n_H<-2$. If, for $n_B<-2$,  $C^{\Theta \Theta}_{(A)l}/C^{\Theta
\Theta}_{(S)l}$ is $l$ independent, then $n_H \simeq n_B$. If $n_B>-2$,  
  $C^{\Theta \Theta}_{(A)l}/C^{\Theta
\Theta}_{(S)l}$ is $l$ independent for any allowed $n_H \geq n_B >-2$.

It is possible that there are other ways to detect
magnetic helicity. On the other hand,  Ref.~\cite{ensslin03} argues that
a detection of magnetic helicity,  even for
cluster magnetic fields, is a very difficult task.

\acknowledgments
We thank Chiara Caprini, Ruth Durrer, and Arthur Kosowsky
for  fruitful discussions and suggestions.
We  acknowledge helpful comments from
Karsten Jedamzik, Antony Lewis, Andy Mack, Tanmay Vachaspati,
and Larry Weaver.
T.K. thanks Geneva University for hospitality
when this work was started. We  acknowledge support
from CRDF-GRDF grant 3316, NSF CAREER grant AST-9875031, and
DOE EPSCoR grant DE-FG02-00ER45824.
\appendix
\section{Derivation of magnetic field source terms for
vorticity perturbations} \label{sec:tau} The derivation of the
symmetric magnetic field source term for the vorticity
perturbation equation of motion is given in \cite{mack02}. To
obtain the complete magnetic field source term for the vector
metric perturbation we use the Fourier transformed magnetic field
energy-momentum tensor, 
\be \tau_{ij}({\bf k}) = {1 \over 2
(2\pi)^4} \int d^3\!p ~\left[B_i({\bf p}) B_j({\bf k-p}) -
{1\over 2}B_l({\bf p})B_l({\bf k-p})\delta_{ij} \right]~,
\label{taub} \ee and its wavenumber space power spectrum $\langle
\tau^{\star}_{ij}({\mathbf k}) \tau_{lm}({\mathbf k^\prime})
\rangle$ \cite{mack02}. 
The symmetric part of the magnetic source power spectrum 
$f(k)$ and the helical part $g(k)$ of the magnetic 
source power spectrum 
can be obtained 
via Eqs.~(\ref{vector-source-sym1}) and (\ref{vector-source-hel}), 
 see Eqs.~(A4)--(A7) of \cite{mack02} 
 and Eq.~(A1) of Ref.~\cite{caprini03}.  
 The part of $f(k)=f_B(k)-f_H(k)$ that depends on  the symmetric
part of the magnetic field power spectrum,  $f_B(k) \sim \int
d^3\!p\,P_B(p)P_B(|\mathbf {k-p}|)$,  is \cite{mack02},
\begin{equation}
f_B(k) \simeq  \frac{\lambda^3 B^4_\lambda }
{16 (2n_B+3)\Gamma^2(n_B+3/2) }
\left[
(\lambda k_D)^{2n_B+3}+\frac{n_B}{n_B+3}(\lambda k)^{2n_B+3}
\right].
\label{vector-symmetric-appendix}
\end{equation}
The magnetic helicity contribution,
$f_H(k) \sim \int d^3\!p\,P_H(p) P_H(|{\bf k}- {\bf p}|)$,
to $f(k)$ is, 
\be
f_H(k) = {1 \over 8 (2\pi)^5}
\int d^3\!p~P_H(p)P_H(|\mathbf {k-p}|)
{{p (1-\gamma^2)}
\over {\sqrt{k^2-2kp\gamma +p^2}}}, 
\label{vector-appendix}
\ee
where $\gamma = {\mathbf{\hat k}}\cdot {\mathbf{\hat p}}$. 
The function $f_H(k)$ is positive but it contributes to $f(k)$ with a
minus sign, $f(k)=f_B(k)-f_H(k)$, and so it
decreases the overall symmetric magnetic field source term.
This result contradicts that of Ref.~\cite{pogosian02}.
Reality requires $P_B(k) \geq |P_H(k)|$,
and the total symmetric source term is always positive, $f(k)>0$.
The mixed terms $\int d^3\!p ~P_B(p)P_H(|\mathbf {k-p}|)$
and
$\int d^3\!p ~P_H(p)P_B(|\mathbf {k-p}|)$ do not contribute to $f(k)$, while
the helical spectrum $g(k)$ is completely defined
by these terms,
\bea
g(k) = {1 \over 16 (2\pi)^5} \int d^3\!p~
P_B(p)P_H(|\mathbf {k-p}|)
{{(k-2p\gamma) (1-\gamma^2)}
\over {\sqrt{k^2-2kp\gamma +p^2}}}.
 \label{vector-appendix1}
\eea
The helical spectrum $g(k)$ does not receive a  contribution from
the diagonal terms $\int d^3\!p ~P_B(p)P_B(|\mathbf {k-p}|)$ and
$\int d^3\!p ~P_H(p)P_H(|\mathbf {k-p}|)$.

To evaluate the expressions for $f_H(k)$ and $g(k)$, we
first integrate over $\gamma$ and then integrate  over $p$.
The integration over $\gamma$ can be done analytically,
see the Appendix of  Ref.~\cite{mack02}.
To integrate over $p$ we approximate the result of
the $\gamma$ integration by using the binomial expansion
$(1+x)^n=1+nx + n(n-1)
x^2/2 + \mathcal{O}(x^3)$.
For the vector case the dominant terms for $x \ll 1$
are those of quadratic order.
Additionally,  the integration  is split into
two parts, $\int_0^{k_D} dp = 
\int_0^{k} dp  + \int_{k}^{k_D} dp $, and
the binomial expansion for $k>p$  used for
$\int_0^{k} dp$, while the second integral  $\int_k^{k_D} dp$ 
 is evaluated
using the binomial expansion  for $k<p$ \cite{mack02,caprini03}.
The result is the vector perturbation source power spectrum
symmetric ($f(k)$) and helical ($g(k)$) terms in
Eqs.~(\ref{V-S-Source}) and (\ref{V-A-Source}) above.

\section{Bessel Functions Integrals}
We need to evaluate integrals of the form
$\int^{x_S}_0 d x\,J_p(ax)J_q(ax)x^{-b}$,
 which contain products of Bessel functions. For
$b >0$ when the integral converges and is dominated by $x\ll x_S$,
the upper limit $x_S$ can be replaced by $\infty$ (with an accuracy of a
few percent for $b>1$, and 15--30 \% for $0<b<1$, depending on the value of
$p-q$). We can then use
Eq.~(6.574.2) of  Ref.~\cite{gradshteyn94},
\be
\int^{\infty}_0dx\,J_p(ax)J_q(ax)x^{-b}=
\frac{a^{b-1}\Gamma(b)\Gamma((p+q-b+1)/{2})}
{2^b\Gamma((-p+q+b+1)/2)
\Gamma((p+q+b+1)/{2})
\Gamma((p-q+b+1)/{2})},
\label{eq:GR-6.574.2}
\ee
which is valid for $\text{Re}\,(p+q+1)>\text{Re}\,b>0$, and $a>0$.

To evaluate the integral $\int^{x_S}_0dx\,x^pJ^2_{l+1/2}(x)$
with $p>0$ and $x_S \gg l$, we use the
asymptotic expansion of $J_p(x)$ for large arguments, Eq.~(9.2.1) of
Ref.~\cite{abramowitz72},
$J_{l+1/2}(x)\simeq\sqrt{2/(\pi x)}\cos[x-(l+1)\pi/2] \simeq
\sqrt{2/(\pi x)}\cos[x-(l+1)\pi/2]$. Replacing the oscillatory
function $\cos^2$ by its r.m.s. value of $1/2$,
we obtain \cite{mack02,caprini03},
\begin{equation}
\pi \int^{x_S}_0dx\,x^pJ^2_{l+1/2}(x)
\simeq\pi \int^{x_S}_{l}dx\,x^pJ^2_{l+1/2}(x)
\simeq
\cases{x^p_S/p,                     & $p>0$\cr
       \ln(x_S/(l+1/2)),            & $p=0$~.\cr}
\label{bessel-int}
\end{equation}

For the integral $\int_0^{x_S} dx\, x^p J_{l+1/2}(x)J_{l+3/2}(x)$
we also use the large argument ($x\gg l$)
approximation for the Bessel functions,  and  find (see Ref.~\cite{caprini03}
for a numerical check),
\begin{eqnarray}
\pi xJ_{l+1/2}(x)J_{l+3/2}(x) &\simeq& {2}\cos\left(x-(l+1){\pi\over 2}
\right)\cos\left(x-(l+2){\pi\over 2}\right)
= {2}\cos\left(x-(l+1){\pi\over 2}\right)
\sin\left(x-(l+1){\pi\over 2}\right)=
\nonumber
\\
&=&\sin \left(2x-(l+1)\pi\right) = {(-1)^{l+1}}\sin(2x).\label{jjj}
\end{eqnarray}
So for $p>0$,
\begin{equation}
\pi\int_0^{x_S}dx\,x^pJ_{l+1/2}(x)J_{l+3/2}(x)\simeq
{(-1)^{l+1}}\int^{x_S}_{l}dx\,x^{p-1}\sin\left(2x\right)
\simeq {(-1)^{l} \over 2}\left(x_S^{p-1}\sin(2x_D)-
(l+1/2)^{p-1}\sin(2l)\right) ~.
\label{bessel-int1}
\end{equation}
This approximation tends to underestimate; it is good to a few percent for
$p>1$ and is within 30 \% for $0 < p\leq 1$.
For  $p=0$ the integral
 $\int_0^{x_S} dx\, J_{l+1/2}(x)J_{l+3/2}(x)$ may be evaluated
using Eq.~(11.4.42) of Ref.~\cite{abramowitz72},
\begin{equation}
\int_0^{x_S} dx\, J_{l+1/2}(x)J_{l+3/2}(x)=\frac{1}{2}.
\label{bessel-int2}
\end{equation}

For the integral $\int_0^{x_S} dx\, x^p [(l+1)J_{l+1/2}(x)/x -
J_{l+3/2}(x)]^2$ appearing in Eq.~(\ref{V-E-power-spectrum}),
for large enough $l$ we have the approximation,
\be
\left(\frac{l+1}{x}J_{l+1/2}(x)-J_{l+3/2}(x)\right)^2
\approx [J^{\prime}_{l+1/2} (x)]^2=
\frac{1}{4}[J_{l-1/2}(x)-J_{l+3/2}(x)]^2.
\label{bessel-approx}
\ee
We now  approximate
the cross term $J_{l-1/2}(x)J_{l+3/2}(x)$ in the limit
$x\gg l$ by using,
\be
\pi x J_{l-1/2}(x)J_{l+3/2}(x) \simeq
{2}\cos\left(x - l{\pi\over 2}
\right)\cos\left(x-(l+2){\pi\over 2}\right)
=-2\cos^2\left(x-l\frac{\pi}{2}\right).
\label{bessel-approx1}
\ee
As in the computation of Eq.~(\ref{bessel-int}) we may replace
the $\cos^2$ by $1/2$, and so find for $p \geq 0$,
\bea
\pi \int_0^{x_S}dx\,x^p \left(\frac{l+1}{x}J_{l+1/2}(x)-J_{l+3/2}(x)\right)^2
\simeq
\frac{\pi }{4}\int_0^{x_S}dx\,x^p
\left(J_{l-1/2}^2 - 2J_{l-1/2}J_{l+3/2} +J_{l+3/2}^2\right)
\nonumber\\
\simeq \pi \int_0^{x_S}dx\,x^p J^2_l
\simeq
\cases{x^p_S/p,                     & $p>0$\cr
       \ln\left({x_S}/(l+1/2)\right), & $p=0$~.\cr}
\label{bessel-int3}
\eea


\begin{thebibliography}{}
\bibitem{lifshitz}
E.~M.~Lifshitz, Zh.~Eksp.~Teor.~Fiz. {\bf 16}, 587 (1946) [English translation:
J. \ Phys.  {\bf 10}, 116 (1946)];
~I.~D.~Novikov, Astron.~Zh.~{\bf 41}, 1075 (1964),
[English translation: Sov.~Astron.
{\bf 7}, 587 (1964)];~J.~M.~Bardeen, Phys. \ Rev. \ D {\bf 22},
1882 (1980); P.~J.~E.~Peebles, {\it
The Large-Scale Structure of the Universe}, (Princeton University, Princeton, 1980),~Sec. V;
~B.~Ratra, \ Phys. \ Rev. \ D
{\bf 38}, 2399 (1988);
V.~F.~Mukhanov, H.~A.~Feldman, and R.~H.~Brandenberger, Phys. \ Rept.
{\bf 215}, 203 (1992);
R.~Durrer, Fund. \ Cosmic \ Phys. {\bf 15}, 209 (1994).


\bibitem{rehnan}A.~Rebhan, Astrophys. \ J. {\bf 392}, 385 (1992);
~A.~Rebhan and D.~J.~Schwarz, Phys. \ Rev. \ D. {\bf 50}, 2541 (1994);~A.~Lewis, Phys. \ Rev. \ D {\bf 70}, 043518 (2004).



\bibitem{adams} J.~Adams, U.~H.~Danielsson, D.~Grasso, and
H.~Rubinstein,  Phys.\ Lett.\ B {\bf 388}, 253 (1996).

\bibitem{durrer98} R.~Durrer, T.~Kahniashvili, and A.~Yates,
Phys.\ Rev.\ D {\bf 58}, 123004 (1998).

\bibitem{subramanian98b} K.~Subramanian and J.~D.~Barrow,
Phys.\ Rev.\ Lett.\ {\bf 81}, 3575 (1998); T.~R.~Seshadri and K.~Subramanian,
Phys.\ Rev.\ Lett.\ {\bf 87}, 101301 (2001);
K.~Subramanian, T.~R.~Seshadri, and  J.~D.~Barrow,
Mon.~Not.~Roy.~Astron.~Soc. {\bf 344}, L31 (2003).

\bibitem{mack02} A.~Mack, T.~Kahniashvili, and A.~Kosowsky,
Phys.\ Rev.\ D  {\bf 65}, 123004 (2002).


\bibitem{lewis04b} A.~Lewis, Phys. \ Rev. \ D. {\bf 70}, 043011 (2004).

\bibitem{review2} For reviews see 
L.~M.~Widrow, Rev.\ Mod.\ Phys. {\bf 74}, 775 (2002);~M.~Giovannini, Int.\ J.\ Mod.\ Phys.\ D {\bf 13},
391 (2004);
~J.~P.~Vall{\'e}e, New~Astron.~Rev. {\bf 48}, 763 (2004).

\bibitem{review3} B.~Ratra, \ Astrophys. \ J. \ Lett. {\bf 391}, L1 (1992);
K.~Bamba and J.~Yokoyama, \ Phys. \ Rev. \ D {\bf 69}, 043507 (2004). 


\bibitem{pogosian02}  L.~Pogosian, T.~Vachaspati, and S.~Winitzki,
Phys.\ Rev.\ D {\bf 65}, 3264 (2002).

\bibitem{gang04} G.~Chen, P.~ Mukherjee, T.~Kahniashvili, B.~Ratra,
and Y.~Wang, Astrophys. J. {\bf 611}, 655 (2004); P.~D.~Naselsky,
L.-Y.~Chiang, P.~Olesen, and O.~V.~Verkhodanov,
\ Astrophys. \ J. {\bf 615}, 45 (2004).

\bibitem{adams96}K.~Subramanian and J.~D.~Barrow,
Phys.\ Rev.\ D {\bf 58}, 083502 (1998);
R.~Durrer, P.~G.~Ferreira, and T.~Kahniashvili,
Phys.\ Rev.\ D {\bf 61}, 043001 (2000); S.~Koh and C.~H.~Lee,
Phys. \ Rev. \ D {\bf 62}, 083509 (2000).

\bibitem{jedamzik98} K.~Jedamzik, V.~Katalini\'c, and A.~V.~Olinto,
Phys.\ Rev.\ D {\bf 57}, 3264 (1998).


\bibitem{caprini03} C.~Caprini, R.~Durrer, and T.~Kahniashvili,
\ Phys. \ Rev. \ D {\bf 69}, 063006 (2004).

\bibitem{cornwall97} J.~M.~Cornwall, Phys. \ Rev. \ D {\bf 56},
6146 (1997);~T.~Vachaspati,
Phys.\ Rev.\ Lett. {\bf 87}, 251302 (2001); ~A.~Brandenburg and E.~Blackman,
astro-ph/0212019, IAU Symp. {\bf 210}, 233 (2003).

\bibitem{jedamzik04} R.~Banerjee and K.~Jedamzik,
Phys.\ Rev.\ D {\bf 70}, 123003 (2004);
V.~B.~Semikoz and D.~D.~Sokoloff, astro-ph/0411496.

\bibitem{kosowsky96} A.~Kosowsky and A.~Loeb,
Astrophys.\ J. {\bf 469}, 1 (1996);~T.~Kolatt,
Astrophys.\ J. {\bf 485}, 564 (1998); S.~Sethi, Mon. \ Not.
\ Astron. \ Soc. {\bf 342}, 962 (2003).

\bibitem{ensslin03} T.~Ensslin and C.~Vogt, Astron.\ Astrophys. {\bf 401},
835 (2003).

\bibitem{campanelli04} L.~Campanelli, A.~D.~Dolgov, M.~Giannotti, and
F.~L.~Villante, \ Astrophys.  \ J. {\bf 616}, 1 (2004).

\bibitem{far} A.~Kosowsky, T.~Kahniashvili, G.~Lavrelashvili, and
B.~Ratra,  Phys. \ Rev. \ D {\bf 71}, 043006 (2005).

\bibitem{scoccola04} E.~Scannapieco and P.~Ferreira,
\ Phys. \ Rev. \ D {\bf 56}, R7493 (1997);
~ C.~Scoccola, D.~Harari, and S.~Mollerach,
\ Phys. \ Rev. \ D {\bf 70}, 063003 (2004).

\bibitem{hu97} W.~Hu and M.~White,
Phys.\ Rev.\ D {\bf 56}, 596 (1997).

\bibitem{durrer03} R.~Durrer and C.~Caprini,
J.~Cosmol.~Astropart.~Phys.  {\bf 11}, 10 (2003).


\bibitem{varshalovich89}  D.~A.~Varshalovich, A.~N.~Moskalev,
and V.~K.~Khersonskii, {\it Quantum Theory
of Angular Momentum}, (World Scientific, Singapore, 1988).

\bibitem{kamionkowski97b} A.~Kosowsky, Ann.~Phys.~{\bf 246}, 49 (1996);
~M.~Zaldarriaga and U.~Seljak, \ Phys. \ Rev. \ D {\bf 55},
1830 (1997);
~M.~Kamionkowski, A.~Kosowsky, and A.~Stebbins,
Phys.\ Rev.\ D {\bf 55}, 7368 (1997).

\bibitem{review4}For reviews  
see K.~Subramanian, astro-ph/0411049;~A.~Challinor astro-ph/0502093.

\bibitem{dolag} K.~Dolag, M.~Bartelmann, and H.~Lesch,
Astron.\ Astrophys.\ {\bf 348}, 351 (1999),~
{\bf 387}, 383 (2002).

\bibitem{caprini00} C.~Caprini and R.~Durrer, Phys.~Rev.~D {\bf 65},  
023517 (2002).

\bibitem{lensing} M.~Zaldarriaga and U.~Seljak,
\ Phys. \ Rev. \ D {\bf 58},
023003 (1997);~A.~Challinor and A.~Lewis, 
astro-ph/0502425. 


\bibitem{gradshteyn94} I.~S.~Gradshteyn and I.~M.~Ryzhik,
{\it Table of Integrals, Series, and Products}
(Academic Press, New York, 1994).

\bibitem{abramowitz72} M.~Abramowitz and I.~Stegun,
{\it Handbook of Mathematical Functions} (Dover, New York, 1972).

\end{thebibliography}
\end{document}